# The most controversial topics in Wikipedia: A multilingual and geographical analysis

**Taha Yasseri[1,2], Anselm Spoerri[3], Mark Graham[1], and János Kertész[4,2]**

[1] *Oxford Internet Institute, University of Oxford, Oxford, United Kingdom*
[2] *Department of Theoretical Physics, Budapest University of Technology and Economics, Budapest, Hungary*
[3] School of Information and Communication, Rutgers University, New Brunswick, New Jersey, USA
[4] *Center for Network Science, Central European University, Budapest, Hungary*

## Abstract

We present, visualize and analyse the similarities and differences between the controversial topics related to "edit wars" identified in 10 different language versions of Wikipedia. After a brief review of the related work we describe the methods developed to locate, measure, and categorize the controversial topics in the different languages. Visualizations of the degree of overlap between the top 100 lists of most controversial articles in different languages and the content related to geographical locations will be presented. We discuss what the presented analysis and visualizations can tell us about the multicultural aspects of Wikipedia and practices of peer-production. Our results indicate that Wikipedia is more than just an encyclopaedia; it is also a window into convergent and divergent social-spatial priorities, interests and preferences.

# 1. Introduction

Value creation in electronic collaborative environments is rapidly gaining importance. Examples include the Open Source Software project, applications for social network services and the paradigmatic case of Wikipedia. The latter is especially suited for scientific research as practically all changes and discussions are recorded and made publicly available. This provides a unique opportunity to study the laws of peer production, the process of self-organization of hierarchical structures needed to make such a system efficient and the occurring regional and cultural differences.

One of the challenges is to understand the emergence and resolution of conflicts in peer production. While the common aim in the collaboration is clear, unavoidably differences in opinions and views occur, leading to controversies. Clearly, there is a positive role of the conflicts: if they can be resolved in a consensus, the resulting product will better reflect the state of the art then without fighting them out. However, there are examples, where no hope for a consensus seems in sight – then the struggle strongly limits efficiency. The investigation of conflicts in Wikipedia can contribute not only to the understanding of the mechanisms of peer production but also to that of the nature of the conflicts itself. What are the dynamics of the emergence of a conflict? What are the main elements of the escalation? How are the camps structured? What are the main techniques of reaching consensus? These are all general questions and there is a large amount of literature about them in political sciences, sociology and social psychology [see, e.g., T.C. Schelling: The Strategy of Conflict (Harvard UP, 1980), Ho-Wen Jong: Understanding Conflict and Conflict Analysis (SAGE, 2008)]. A major problem in the quantitative analysis of conflicts is the lack of appropriate measures and sufficient amount of data. Wikipedia is a special environment but its complete documentation makes it particularly suited for such quantitative studies. We hope to gain information not only about conflicts and their resolution under collaborative task solving circumstances but also about the general nature of controversies. In addition, the presence of different editions of Wikipedia for different languages, allows us to investigate all the mentioned research questions on a global scale and cross-lingually, aiming at understanding universal and local features of the conflicts. Cross-cultural comparisons of contested and controversial topics, provides us with a detailed and naturally generated image of the priorities and sensitivities of the editors' community of the specific language edition.

Our interest in this study is mainly of a socio-political nature. By using and comparing 12 different language editions of Wikipedia, we were interested in cultural differences and similarities and we investigate whether Wikipedia is a multicultural forum or there are strong linguistic-group-dependent features. Here we employ visualizations to assist us in disentangling this multivariate issue. Our findings suggest that on the one hand Wikipedia as a unifying tool brings divergent groups of individuals closer to each other, on the other hand there are still specific characteristics to be understood only based on localities and cultural differences. One approach to this issue is to investigate the effects coming from geographical locations of controversial topics in different language editions of Wikipedia.A further approach is to identify and visualize the contested topics that are shared in different languages and cultures as well as show which topics are specific to a language.

There is an increasing body of literature on Wikipedia conflicts; for a recent review see Yasseri and Kertész (2013). The first problem to solve is to create an automated filter to identify controversial articles that works independently of all the languages of Wikipedia. While there are lists of controversial topics and “Lamest edit wars” in Wikipedia provided by editor communities, it has been argued that those lists are neither complete nor exclusive (Sumi 2011a) As such, there have been several efforts to introduce quantitative algorithms to detect and rank controversial articles in Wikipedia (Kittur et al. 2007, Vuong et al. 2008, Sumi et al. 2011a,b) . Here a compromise between efficiency and accuracy is called for. Among the one-dimensional measures that can be computed, the one developed by Sumi et al. (2012) turns out to be one of the most reliable ones (Sepehri Rad and Barbosa, 2012). Using this measure, detailed studies of the statistics (Sumi et al., 2011), dynamics and characteristics of conflicts in different versions of Wikipedia (Yasseri et al., 2012a) were carried out. In this work, we use the same methodology to locate and rank the controversial articles and then compare their topical coverage and geographical locations (where possible) across different language editions.

Previous work on topical coverage of contested articles in English Wikipedia (Kittur 2009) has reported that “Religion” and “Philosophy” are among the most debated topics. However, this study doesn't give more detailed information about the individual articles with high level of controversy and not a comparison between different language editions either. The detection methods used by Kittur et al. are based on the “Controversial Tags” assigned by editors to the articles, which evidently does not include all the really debated articles and is hard to generalise to language editions beyond English (Sumi et al. 2011b).

Apic et al. (2011) also took the user tagged articles in English Wikipedia into account and calculated a “dispute” index for each country based on the number of tagged articles linking to the article about the country. They show that this index correlates with external measures for governance, political, and economical stability, such that the higher the dispute index is the lower the stability: Wikipedia Dispute Index correlates negatively with the World Bank Policy Research Aggregate Governance Indicator (WGI) for political stability with $R = -0.781$.

## 2. Methods

In this section, we briefly describe the controversy detection methods and the visualization methods used to present the results.

### *2.1 Controversy detection method and topical categorization*

We quantify the controversiality of an article based on its editorial history, by focusing on "reverts", i.e. when an editor undoes another editor's edit completely and brings it to the version exactly the same as the version before the last version. To detect reverts, we first assign a MD5 hash code (Rivest RL, 1992) to each revision of the article and then by comparing the hash codes, detect when two versions in the history line are exactly the same. In this case, the latest edit (leading to the second identical revision) is marked as a revert, and a pair of editors, namely a reverting and a reverted one, are recognized. A "mutual revert" is recognized if a pair of editors *(x, y)* is observed once with *x* and once with *y* as the reverter. The weight of an editor *x* is defined as the number of edits *N* performed by him or her, and the weight of a mutually reverting pair is defined as the minimum of the weights of the two editors. The controversiality *M* of an article is defined by summing the weights of all mutually reverting editor pairs, excluding the topmost pair, and multiplying this number by the total number of editors E involved in the article. This results in the following:

$$M = E \sum_{\text{all mutual reverts}} min(N^{\text{d}}, N^{\text{r}})$$

where N r/d is the number of edits for the article committed by reverting/reverted editor. The sum is taken over mutual reverts rather than single reverts because reverting is very much part of the normal workflow, especially for defending articles from vandalism. The minimum of the two weights is used because conflicts between two senior editors contributing more to controversiality than conflicts between a junior and a senior editor, or between two junior editors. And finally the topmost reverting pair is excluded to avoid overestimating the editorial war dominated by a personal fight between two single editors. The explained measure can be easily calculated for each article, irrespective of the language, size, and length of its history.

## *2.2 Visualization of topical overlaps*

The searchCrystal visualization toolset will be used to compare, visualize and identify Wikipedia pages that are highly contested in multiple languages. Similar to a bullseye display, searchCrystal uses a radial mapping so that the Wikipedia pages contained in *all* the language lists that are being compared are mapped to the *center* of the display and the number of lists that contain the same page decreases toward the periphery of the display. searchCrystal consists of several complementary views: the Category, Cluster, Spiral and List View (Spoerri, 2004a; 2004b; 2004c). Each view helps the user explore specific aspects of the overlap structure between the lists being compared.

Similar to a Venn diagram, the Category View provides an aggregated view since it groups all the pages that are contained in the same combination of lists and shows how many pages are included in which specific combinations of languages (see Figures 2 and 6). At its periphery, star–shaped icons with a single color represent the specific lists that are being compared. Each list is assigned a unique color and the number inside a star–shaped icon indicates the number of pages in the list; the text label next to the star–shaped icon indicates which language (or combination of languages) is used as a crystal input. The interior of the Category View consists of circular icons whose colored sectors indicate which specific lists contain the same page. The size of a circular icon indicates how many pages are contained in a specific combination of lists. At the edge of a circular icon, the two pages with the highest list positions are also shown.

The Cluster View shows how the *individual* Wikipedia pages are related to the lists being compared and a radial mapping is used to map pages into concentric rings based on the number of lists that contain them (see Figures 3, 4 and 5). In this view, the star–shaped icons at the periphery act like "magnets" that pull a page icon toward them based on the page's list positions (Spoerri, 2004a). Thus, the position of a page icon reflects the relative difference between the page's positions in the lists that include it. Further, pages are mapped into the same circular ring if they are contained in the same number of lists. The closer a page is placed toward the display center within a ring, the higher the average of its list positions. A page icon has multiple visual properties to help the user determine how many and which specific lists contain the page and the page's average position in the lists. The shape of a page icon indicates the number of lists that contain the page and the colors indicate which lists. The size of a page icon reflects the average position of the page in the lists that contain it. The greater the size and the stronger the color saturation of a page icon, the higher up it is placed in the lists. Thus, both the position of a page icon inside its designated ring and its size and color saturation indicate whether a page is highly placed in the lists that include it. The page title is displayed next to a page icon, but it can be truncated to prevent titles from overlapping.

The Spiral View places all pages sequentially along an expanding spiral so that distance of page icons from the display center is the same as in the Cluster View (see Figure 7)Pages that are included in all of the lists being compared are located in the center ring. The icons for pages that are contained in the same number of lists are placed consecutively along the spiral and in the same concentric ring. Title fragments are displayed in the radial direction to make effective use of the white space in the spiral layout (Spoerri, 2004b). The Spiral View, which can be rotated, makes it possible for users to rapidly scan a large number of pages and their titles.

### *2.3 Geographic data*

In order to determine the geographic coordinates of Wikipedia articles, we drew upon the efforts of the Oxford Internet Institute's geographic article parser (see Graham 2011 or https://github.com/oxfordinternetinstitute/wikiproject). The data were all taken from 2012 Wikipedia data dumps and we searched for coordinates in every article (if an article had a coordinate in any language edition, we assigned that same coordinate to the equivalent articles in all other languages. We improved the quality of our coordinates by doing things like eliminating or fixing erroneous coordinates and making sure to remove irrelevant coordinates (Wikipedia actually contains a lot of coordinates for extra-terrestrial entities like lunar craters![1]). We also excluded articles that are essentially indexes of places from our geographic dataset as they tell us little about any particular parts of the world (e.g. articles about events, monuments, towns etc.). We thus reduced the dataset to all articles with four or fewer sets of coordinates, and used the coordinates that appear most frequently (if all coordinates appeared once, we used the first set).

1 E.g. http://en.wikipedia.org/wiki/Bailly_%28crater%29

## 3. Results

We have calculated the controversiality *M* for all the articles available in 10 different language editions of Wikipedia based on the data dumps we downloaded in March 2010. We tried to a as diverse as possible sample including West European, East European, and Middle Eastern languages within the language capabilities of our research team. The selected language editions span a wide range in terms of the number of articles (and active editors): English, German, French, with more than 1 million artciles (and 19, 1.7, and 1.6 million editors), Spanish, with more than 500 thousand articles (2.7 million editors), Persian, Czech, Hungarian, Arabic, and Romanian with more than 200 thousand (with 350, 230, 235, 645, and 264 thousand editors), and finally Hebrew with 142 thousand articles and 200 thousand editors. The more than 27 million editors and potential readers of these language editions have a much extended geographical distribution and that enables us to investigate the disputed titles and topics at the global scale.

Different selections of the lists of *M* scores are available at http://wwm.phy.bme.hu/ to download. In Table 1, the top10 lists of the most controversial articles with highest *M*'s are provided. By looking at the titles, already a rough impression on the topical coverage of the editorial wars in each language can be gained. For example, *Jesus* appears in among the top-10 in English, German, French, and Czech Wikipedias. Religion, Politics and Geographical places seem to be the common fields of editorial wars in all editions, however with local effects: far-right politics and nationalism in Hungarian, current Iranian political figures in Persian, Sex and Gender related topics in Czech and football clubs in Spanish Wikipedias are evident examples for these localities. To be able to compare the disputed articles at the title level, we needed to “transform” all the topics to English. To this end, we have used the Wikipedia inter-language links created and modified by editors and automated robots. We replaced all the titles by the corresponding title in the English Wikipedia. However, in a few cases, there was no article about exactly the same topic in English Wikipedia and so we kept the original title. In Figure 1, a world cloud of all the 1000 titles generated by this process is depicted. Size of the words is proportional to their appearance frequency. The Cloud is self-explanatory and it already shows some common patterns among most controversial topics.

We demonstrate and analyse these overlaps at the level of article titles in the next section.

*Table 1* Top-10 most controversial articles in each language edition of Wikipedia. Titles in italic are literally translated; the rest are the titles of the sister articles in English Wikipedia.

| en | De | fr | es | cs |
|---|---|---|---|---|
| George W. Bush | Croatia | Ségolène Royal | Chile | Homosexuality |
| Anarchism | Scientology | Unidentified flying object | Club América | Psychotronics |
| Muhammad | 9/11 conspiracy theories | Jehovah's Witnesses | Opus Dei | Telepathy |
| LWWEe[2] | *Fraternities* | Jesus | Athletic Bilbao | Communism |
| Global warming | Homeopathy | Sigmund Freud | Andrés Manuel López Obrador | Homophobia |
| Circumcision | Adolf Hitler | September 11 attacks | Newell's Old Boys | Jesus |
| United States | Jesus | Muhammad al-Durrah incident | FC Barcelona | Moravia |
| Jesus | Hugo Chávez | Islamophobia | Homeopathy | Sexual orientation change efforts |
| Race and intelligence | Minimum wage | God in Christianity | Augusto Pinochet | Ross Hedvíček |
| Christianity | Rudolf Steiner | Nuclear power debate | Alianza Lima | Israel |

| hu | Ro | ar | fa | he |
|---|---|---|---|---|
| *Gypsy Crime* | FC Universitatea Craiova | Ash'ari | Báb | Chabad |
| Atheism | Mircea Badea | *Ali bin Talal al Jahani* | Fatimah | Chabad messianism |
| *Hungarian radical right* | Disney Channel (Romania) | Muhammad | Mahmoud Ahmadinejad | 2006 Lebanon War |
| Viktor Orbán | Legionnaires' rebellion & Bucharest pogrom | Ali | People's Mujahedin of Iran | B'Tselem |
| *Hungarian Guard Movement* | Lugoj | Egypt | Criticism of the Quran | Benjamin Netanyahu |
| Ferenc Gyurcsány's speech in May 2006 | Vladimir Tismăneanu | Syria | Tabriz | *Jewish settlement in Hebron* |
| *The Mortimer case* | Craiova | Sunni Islam | Ali Khamenei | Daphni Leef |
| *Hungarian Far-right* | Romania | Wahhabi | Ruhollah Khomeini | Gaza War |
| Jobbik | Traian Băsescu | Yasser Al-Habib | Massoud Rajavi | Beitar Jerusalem F.C. |
| Polgár Tamás | Romanian Orthodox Church | Arab people | Muhammad | Ariel Sharon |

2 List of World Wrestling Entertainment, Inc. employees

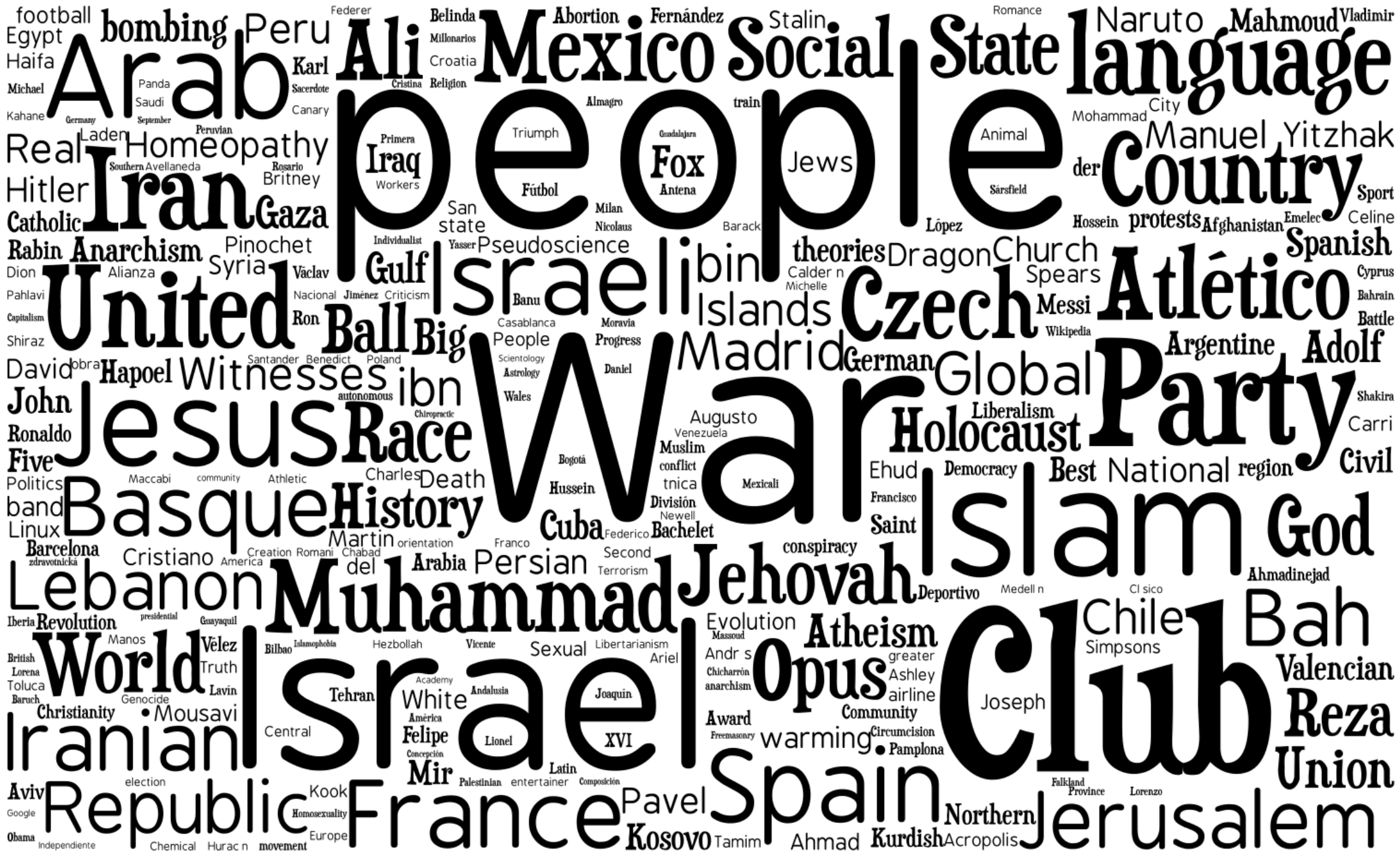

*Figure 1* Word Cloud made of the titles of the 1000 of the most controversial articles in the 10 language editions under study.

### *3.1 Overlapping lists and the patterns*

The searchCrystal visualization toolset is used to first visualize the overlap between these language groupings: 1) English, German, French, Spanish; 2) Czech, Hungarian, Romanian; 3) Arabic, Persian, Hebrew. Next the overlap between these language groupings is visualized to identify Wikipedia pages that are highly contested in multiple language sets. An interactive version of searchCrystal can be accessed at

http://comminfo.rutgers.edu/~aspoerri/searchCrystal/searchCrystal_editWars_ALL.html

and a screencast provides a quick overview how the visualizations were created at

http://comminfo.rutgers.edu/~aspoerri/searchCrystal/WikiEditWars_Screencast/WikiEditWars_Screencast.html.

In Figure 2, the Category View is used to show an aggregated view of the overlap structure between the most contested Wikipedia pages in English, German, French and Spanish: the two pages about *Homeopathy* and *Jesus*, respectively,  are contested in all four languages; s; 5 pages are contested in three out of the four languages and, for example, *Global Warming* and *Socialism* are contested pages in the English, German and French language versions of Wikipedia; 34 pages are contested in two out of four languages. Examining the icons that represent the pages that are contested in only one of the four languages, 67 of the 100 most contested pages in English are not contested in the other three languages.English has the most pages (33%) that are also contested in one or more of the other languages, whereas Spanish has the fewest pages (12%) that are also contested in another language.

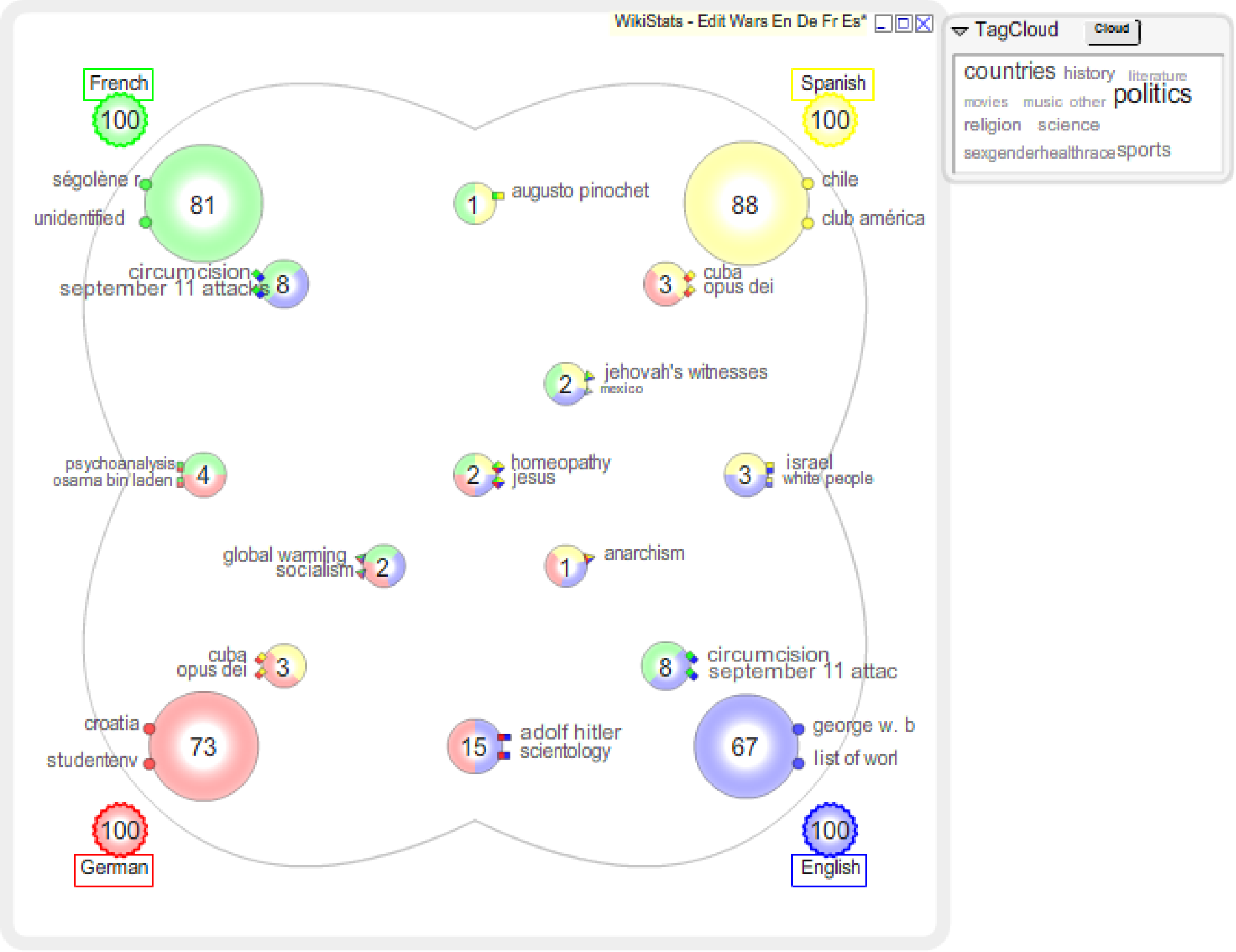

*Figure 2* **Category View** of the overlap structure of the most contested Wikipedia pages in *English*, *German*, *French* and *Spanish*.

In Figure 3, the Cluster View is used to visualize the overlap between the most contested pages in English, German, French and Spanish and it uses a *Fisheye transformation* to visually emphasize the pages that are contested in at least two languages (Spoerri, 2004c). The *Homeopathy* and *Jesus* pages are contested in all the four languages and their diamond shaped icons are placed close to the display center, which indicates that they are on average highly contested in all the four languages. The detail-on-demand display shown below the TagCould panel in the top right corner, shows that the *Jesus* page is one of the top 10 contested pages for English, German and French and the 42nd most contested page for Spanish. For the pages that are contested in three of the four languages, the *Jehovah's Witnesses* page is relatively highly contested in English, French and Spanish, as is the *Anarchism* page in English, German and Spanish and their respective triangular icons are placed close to the display center. The *Global Warming* and *Socialism* pages are not as highly contested in English, German and French and their respective triangular icons are placed close to the middle of the ring that contains pages that are contested in three of the four languages. The *Mexico* page is moderately contested in English, French and Spanish and thus its triangular icon has a smaller size, is less saturated colors and placed further away from the display center than the other pages that are contested in three of the four languages.

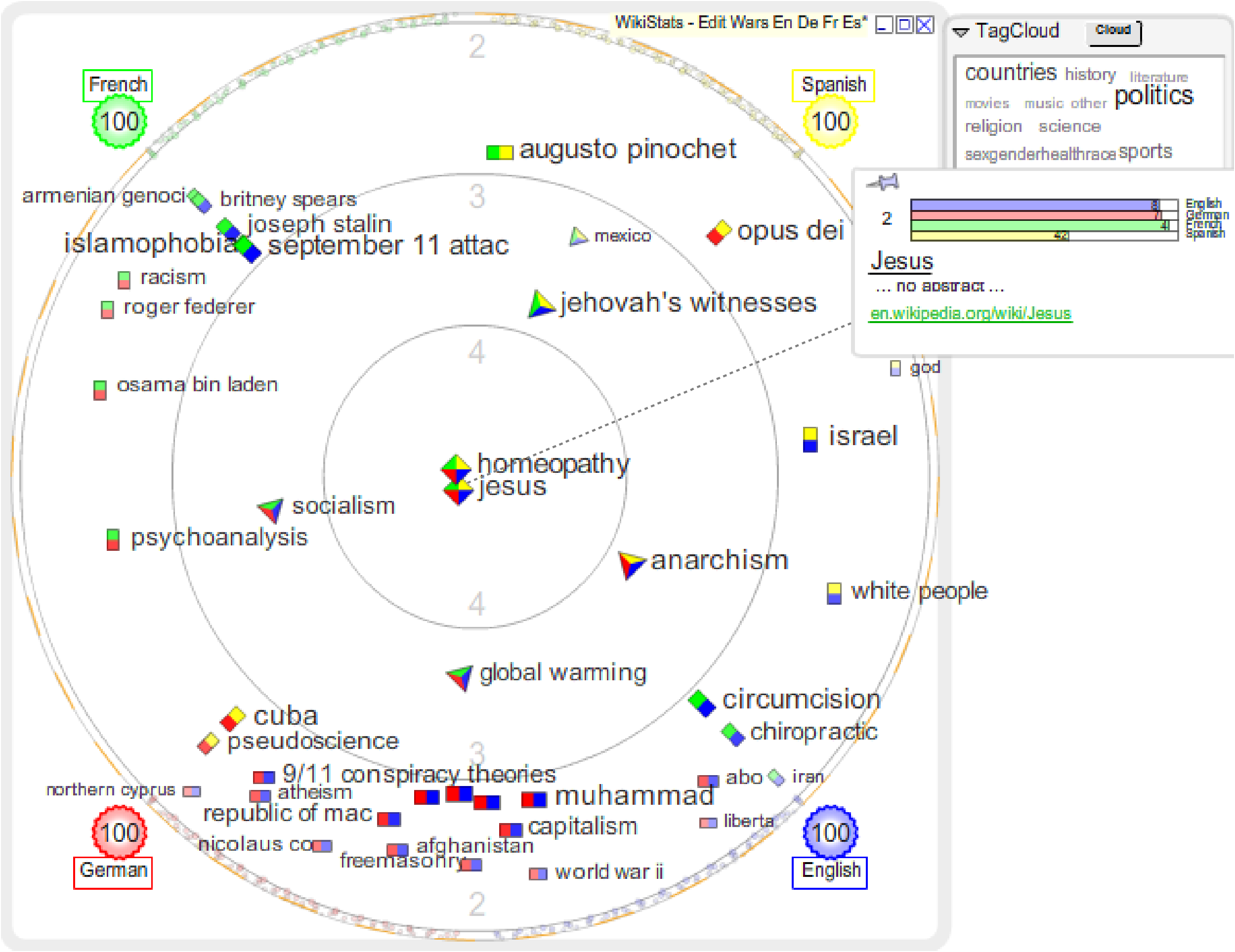

*Figure 3* **Cluster View** of the overlap structure of the most contested Wikipedia pages in *English*, *German*, *French* and *Spanish*.

In Figure 4, the Cluster View is used to visualize the overlap between the most contested pages in Czech, Hungarian, and Romanian and it also uses a Fisheye transformation to visually emphasize the pages that are contested in at least two languages. Specifically, it shows that the Wikipedia page about *Google* is contested in Czech, Hungarian, and Romanian, but its triangular icon is placed away from the center and this indicates that this page is not one of the most contested pages in the respective languages. Further, 5 pages are contested in at least two of the three languages: *The Holocaust* and *Romani People* pages are highly contested in both Hungarian and Romanian since their icons are placed closer toward the center; the *Jesus* page is contested in Czech and Romanian, but more so in the Czech Wikipedia since its icon is placed closer to the star-shaped icon that represents the Czech input list of most contested pages.

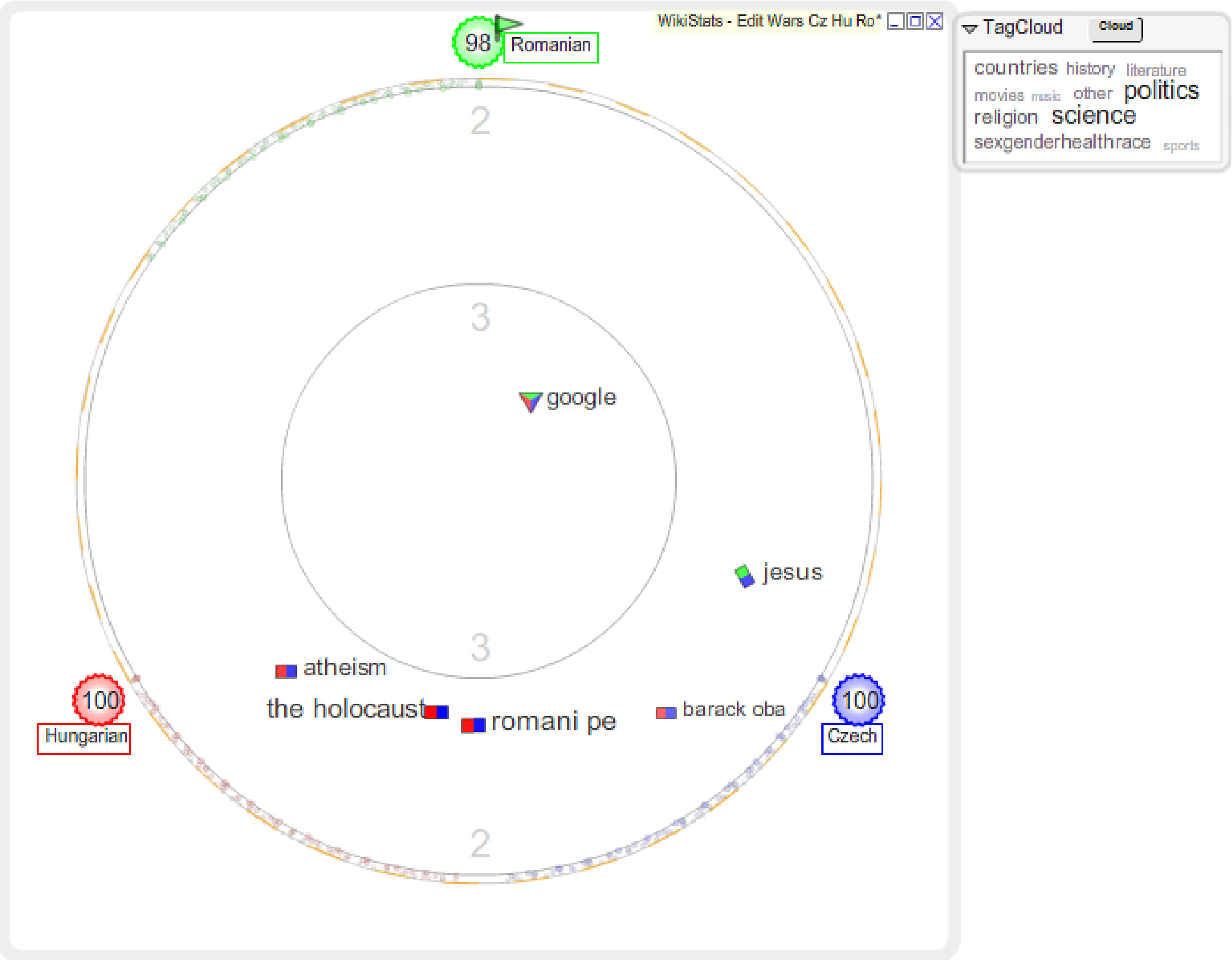

*Figure 4* **Cluster View** of the overlap structure of the most contested Wikipedia pages in *Czech, Hungarian and Romanian* (the flag next to the input icon indicates that there are two Romanian Wikipedia pages with URLs that are identical to other pages).

In Figure 5, the Cluster View visualizes the overlap between the most contested pages in Arabic, Persian, and Hebrew and it also uses a Fisheye transformation to visually emphasize the pages that are contested in at least two languages. 3 pages are contested in all three languages and the *Gaza War* page is the most contested page and its triangular icon is placed quite close to the display center, which indicates that this page is quite highly contested overall, and since its icon is placed closer to the Hebrew and Persian (Farsi) star-shaped input icons and further away from the Arabic input icon, this indicates that the Gaza War page is most contested in Hebrew, a little less so in Persian and less so in Arabic.

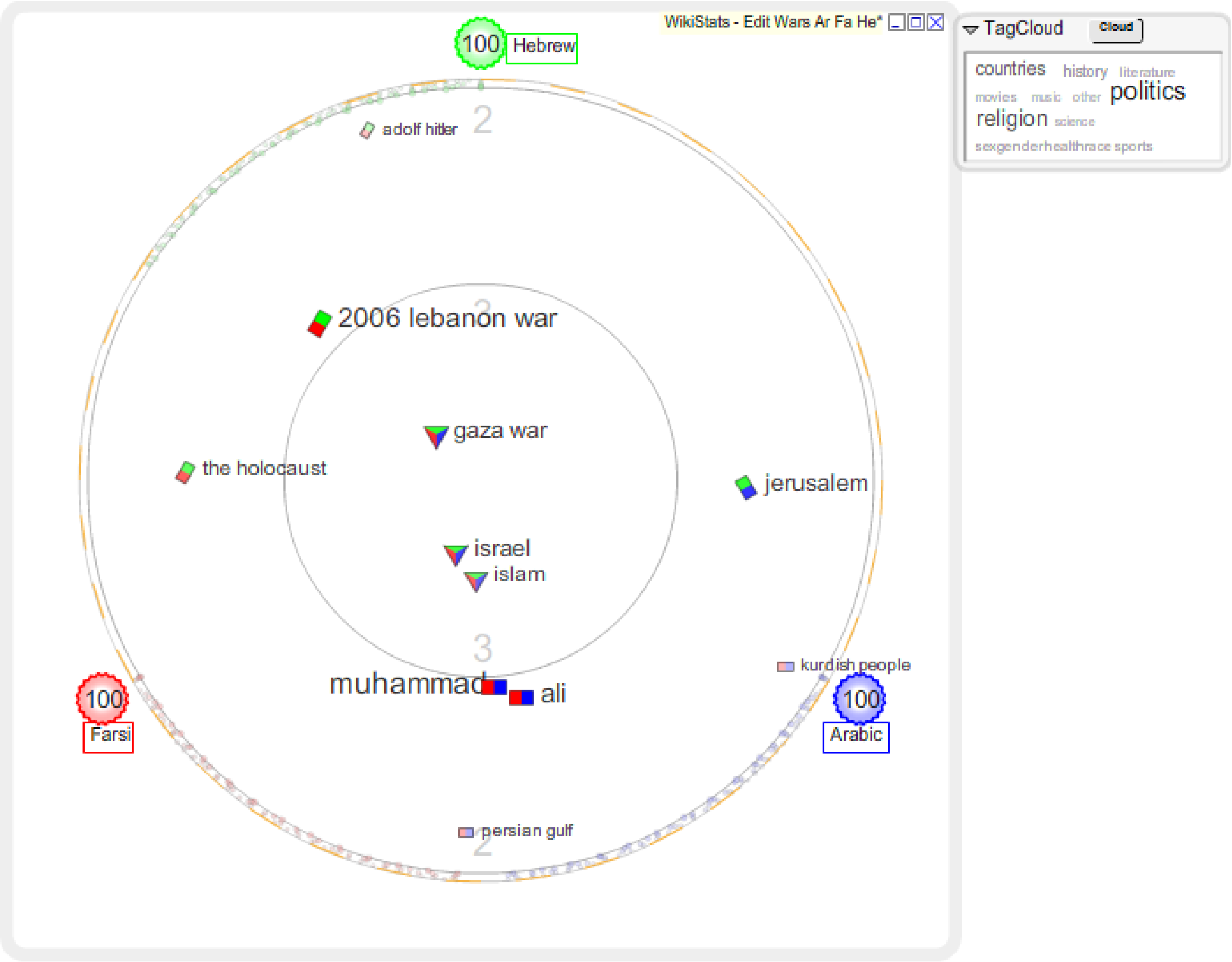

*Figure 5* **Cluster View** of the overlap structure of the most contested Wikipedia pages in *Arabic, Persian* and *Hebrew.*

Altogether 8 pages are contested in two of the Arabic, Persian and Hebrew language versions of Wikipedia: the *Muhammad* page is very highly contested in both Arabic and Persian; the *2006 Lebanon War* page is highly contested in both Hebrew and Persian. Located in the top right corner of the overlap display, a TagCloud panel shows the relative frequency of the categorical topics that are most contested for Arabic, Persian and Hebrew, *political* and *religious* topics represent the two most contested topics.

As a next step in the visual analysis of the most contested pages in the different language versions of Wikipedia, the unique pages contained in the three language sets can be compared with each other. searchCrystal makes it possible to use drag & drop operations – see screencast for demonstration of how this can be done – to compute the overlap structure between the set of English, German, French and Spanish pages (350 unique pages), the set of Czech, Hungarian and Romanian pages (291 unique pages) and the set of Arabic, Persian and Hebrew pages (286 unique pages), as shown using the Category View in Figure 6. There are 7 pages that are contested in all three language sets, where the *Israel* and *Adolf Hitler* pages are the most highly contested pages that are contained in all three language sets. 41 pages are contested in two of the three language sets, where the *Jesus* and *Jehovah's Witnesses* pages are the most highly contested pages that are both contained in the English, German, French Spanish language set and the Czech, Hungarian, Romanian language set.

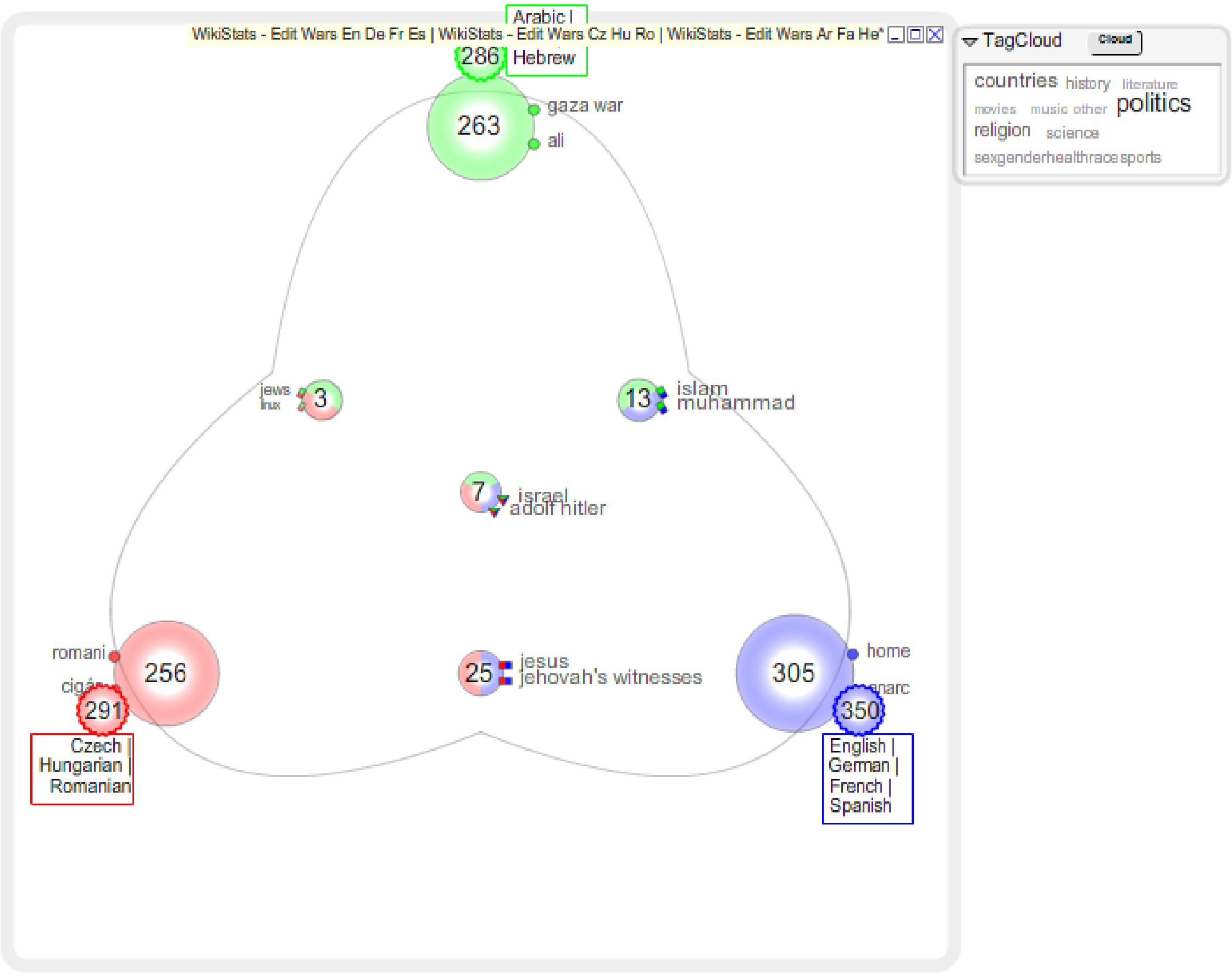

*Figure 6* **Category View** of the overlap structure of the most contested Wikipedia pages in the three sets of languages (blue = English, German, French, Spanish; red = Czech, Hungarian, Romanian; green = Arabic, Persian, Hebrew). As shown in the star-shaped input icons, there are350 unique English, German, French and Spanish pages; 291 unique Czech, Hungarian and Romanian pages; and 286 unique Arabic, Persian and Hebrew pages.

The *Islam* and *Muhammad* pages are the most highly contested pages that are both contained the English, German, French Spanish language set and the Arabic, Persian, Hebrew language set.  Roughly 10% of the pages in each language set are also contested in another language set. The TagCloud panel shows that pages related to *political* topics represent the largest group of contested pages in the union of all three language sets, followed by pages related to *geographical / countries* and *religious* topics, respectively.

In Figure 7, the Spiral View visualizes the overlap structure of the individual pages contained in the set of English, German, French and Spanish pages, the set of Czech, Hungarian and Romanian and the set of Arabic, Persian and Hebrew pages.

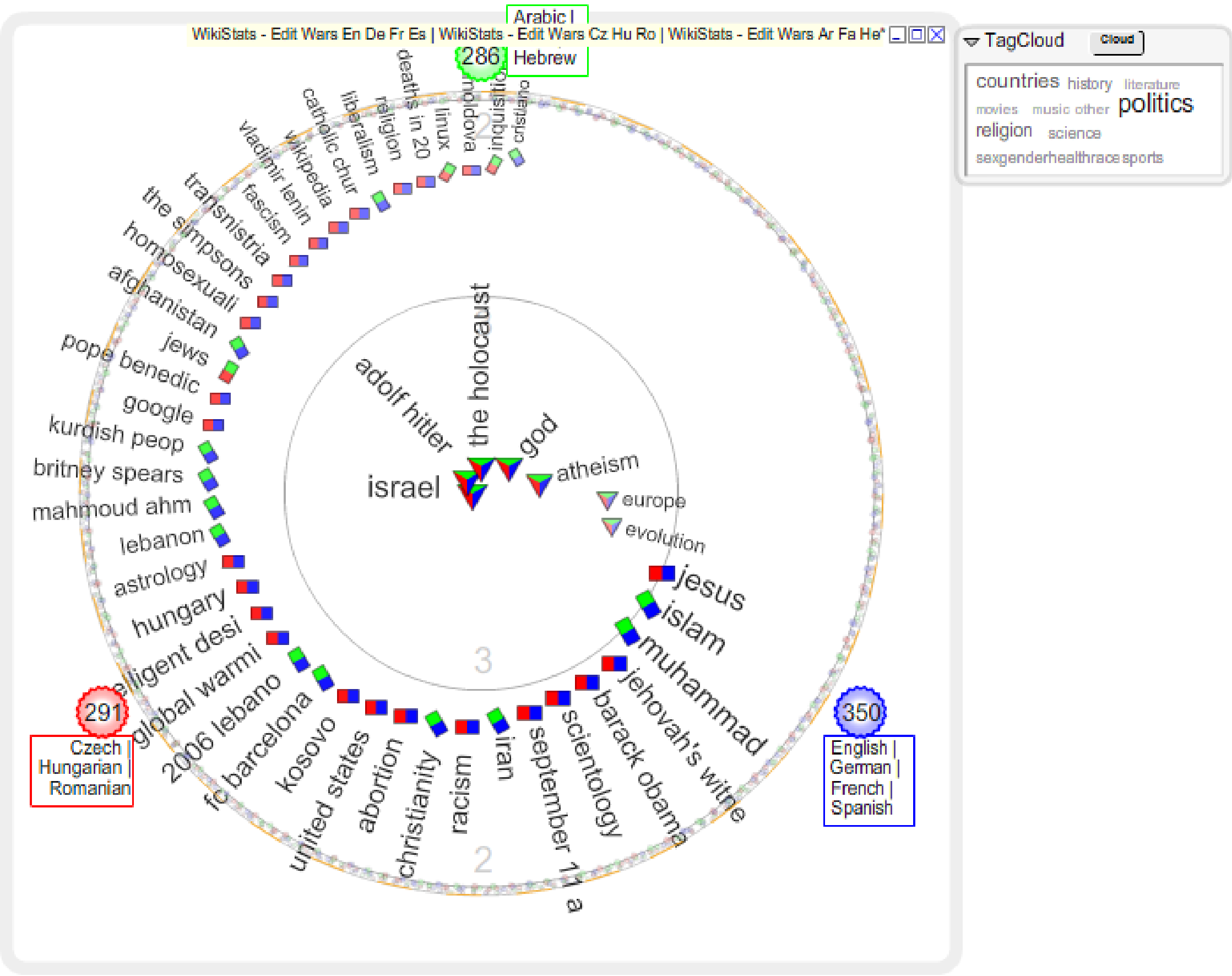

*Figure 7* **Spiral View** of the overlap structure of the most contested Wikipedia pages in the three sets of languages (blue = English, German, French, Spanish; red = Czech, Hungarian, Romanian; green = Arabic, Persian, Hebrew).

A Fisheye transformation is used to visually emphasize the pages that are contained in at least two language sets. As described in more detail in Section 2.2, the closer a page icon is placed toward the display center within a ring, the higher the average of its list positions; the greater the size and the stronger the color saturation of a page icon, the higher up it is placed in the lists. The Spiral View can be rotated so that the titles can be read more easily and so the angle of an icon with respect to the display center does not encode the relative difference between the list positions in the sets being compared as is the case in the Cluster View. The Spiral View is designed to make it easy to see the ranking of the pages in terms of their average list positions and the number of language sets that contain them as well as to ascertain whether the pages contained in more than one language set are highly contested pages are not. The *Israel*, *Adolf Hitler*, *The Holocaust*, *God*, *Atheism*, *Europ*e, *Evolution* pages are contained in all three language sets, and the pages *Israel*, *Adolf Hitler*, *The Holocaust* and *God* are highly contested in all the language sets since their triangular icons are placed close to the center of the display and their sizes and color saturations are close to the highest possible values. The *Jesus*, *Islam*, and *Muhammad* pages are very highly contested in two of three sets of languages since their icons are placed closest toward the display center within the ring that contains pages that are contained in two of the language sets.

### *3.3 Topical Coverage of conflict*

Whilst comparing the titles to locate the overlaps, one realises that there are many articles with slight differences in the title but very similar contents in two or more language editions. Our visualisation method is not capable of capturing these similarities. Therefore we have categorised all the articles in top-100 lists into 10 categories based on their primary categorical tags in Wikipedia and also human judgements about their contents[3]. The 10 categories and the topics each of them includes are described in Table 2. Although at the first glance, these 10 categories might look insufficient to cover all the 1000 categorised articles (100 most controversial articles in 10 languages), but interestingly, there have been only 26 articles out of 1000, which did not

3 Each article is coded by two coders with interceder reliability of 95 to 100% for different languages. In cases of mismatching categories, the decision of the more confident coder (native speaker) has been adopted.

match any of these 10 categories. This indicates a high level of similarities in different language lists at the topical level.

The aggregated populations of the categories for all the 1000 articles are depicted in Figure 8. Politics-related articles exceed one quarter of the whole population and in addition to "geographical places" and "Religion" cover more than half of the most controversial articles. Culture-related articles including, literature, authors, printed and public media, movies and animations, entertainment and music industry, although ranked 8-10 according to the relative population. However putting all these categories together, it goes beyond 10% of the sample.

*Table 2* The topical categories for the controversial articles.

| Category | Topics | Name | % |
|---|---|---|---|
| A | Politics, Politicians, Political Parties, Political Movements and Ideologies | Politics | 25 |
| B | Geographical locations, Countries, Cities, Towns | Countries | 17 |
| C | Religion, Cults, Beliefs | Religion | 15 |
| D | History, Historical Figures | History | 9 |
| E | Sex and Gender, Health, Human Rights, Environment, Social Activism | SexGender-HealthRace | 7 |
| F | Science, Technology, Internet, Web | Science | 7 |
| G | Sport Clubs, Sport People, Sport Events | Sports | 6 |
| H | Literature, Journals, Journalists, Authors, News Papers, Languages | Literature | 4 |
| I | Movies, TV Channels, TV Series, Theatres, Actor, Directors, Animations | Movies | 4 |
| J | Songs, Singers, Music Genres, Music Events | Music | 3 |

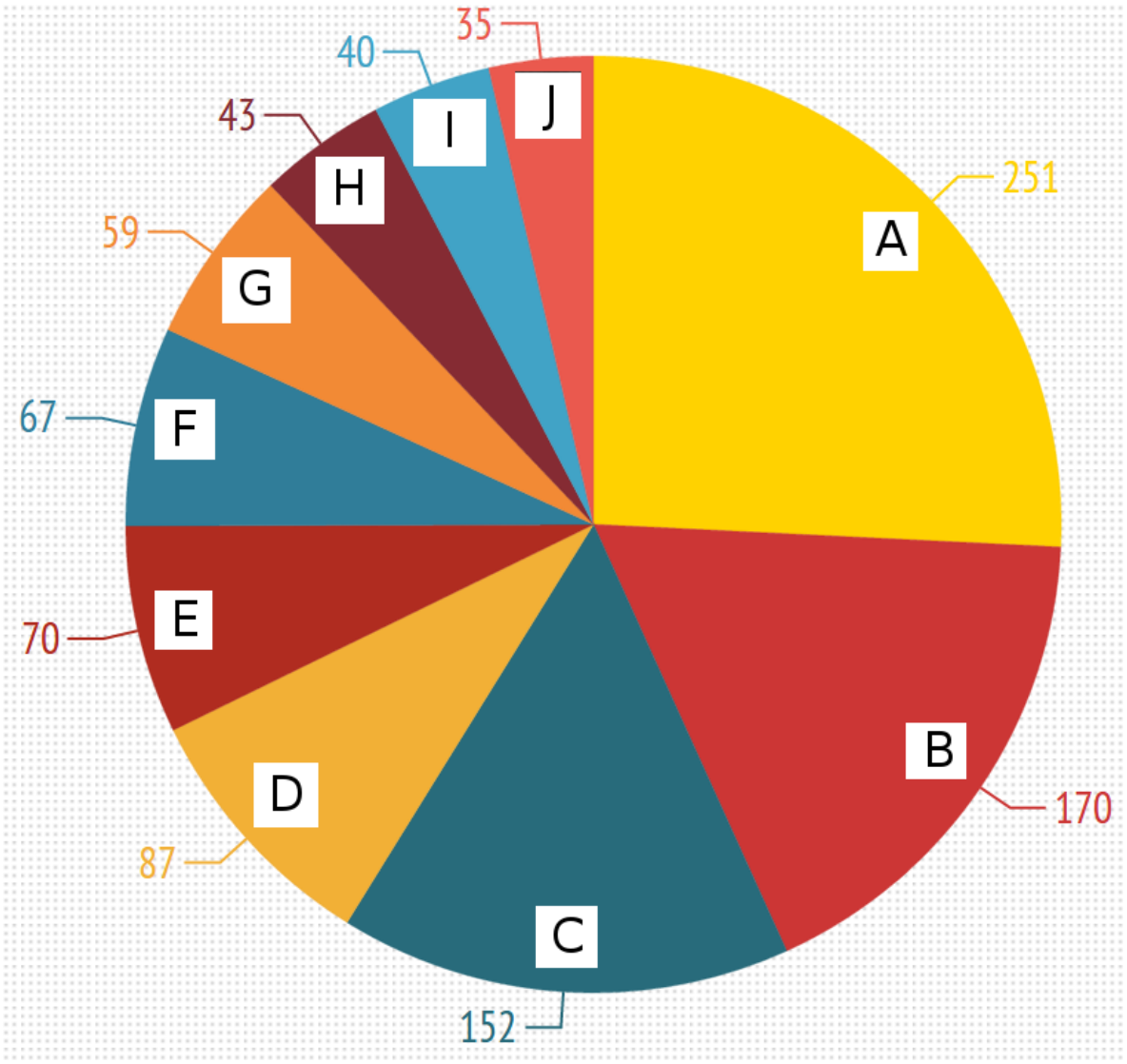

Figure 8  Population of the topical categories of the 1000 most controversial articles in 10 different language editions. Categories are described in Table 2.

Although this pattern is universally the same in all language editions to some good extent, but there are also interesting local deviations from the overall norm in each edition. In Figure 9, populations of the topical categories are shown for each language editions. The predominant examples of anomalies are 1) Category G (sport) in Spanish Wikipedia, 2) Religion, Geographical places and History in Arabic, Persian and Hebrew, 3) Science and Technology related topics in French and Czech and finally 4) Music and Entertainment in Romanian Wikipedia. It is clearly due to cultural differences and the variation of community priorities from one language to others.

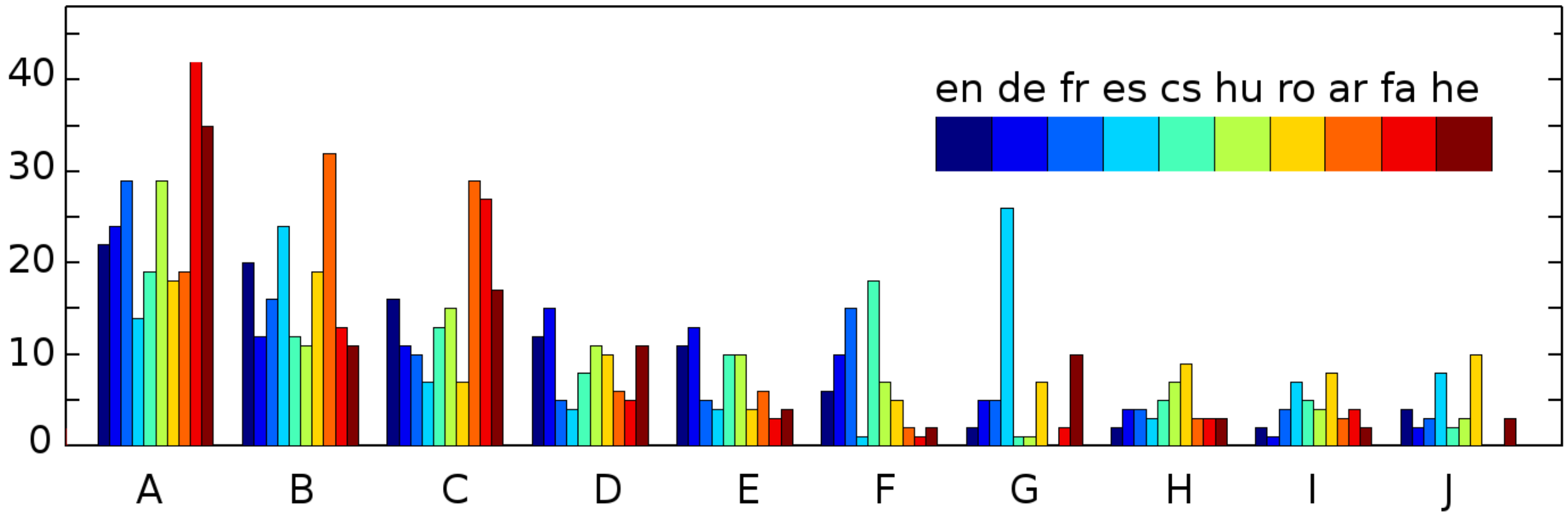

*Figure 9* Category population separated for each language edition, showing the deviations from the universal topical patterns.

### *3.3 Geographical distribution of conflict*

As mentioned earlier, articles on geographical places, countries, cities, towns, etc, are among the most populated conflict categories. Moreover, many of articles which are not directly about places are "geotagged" and therefore are linked to a geographical location. Therefore it could be enormously helpful to make geographical maps of the controversial articles, based on their geotags.

When mapping the geographic dispersion of conflict, we see an interesting amount of difference between the different language versions of Wikipedia. Some of the smaller language Wikipedias have a high-degree of self-focus in their articles that are characterized by the greatest degree of conflict (see also the work of Hecht 2009 for another illustration of how different language communities on Wikipedia tend to write about places close to home). Note, for instance, the geographic focus of conflict in the Czech and Hebrew Wikipedias in Figures 10 and 11 (with the top-five locations of conflict labeled on both maps).

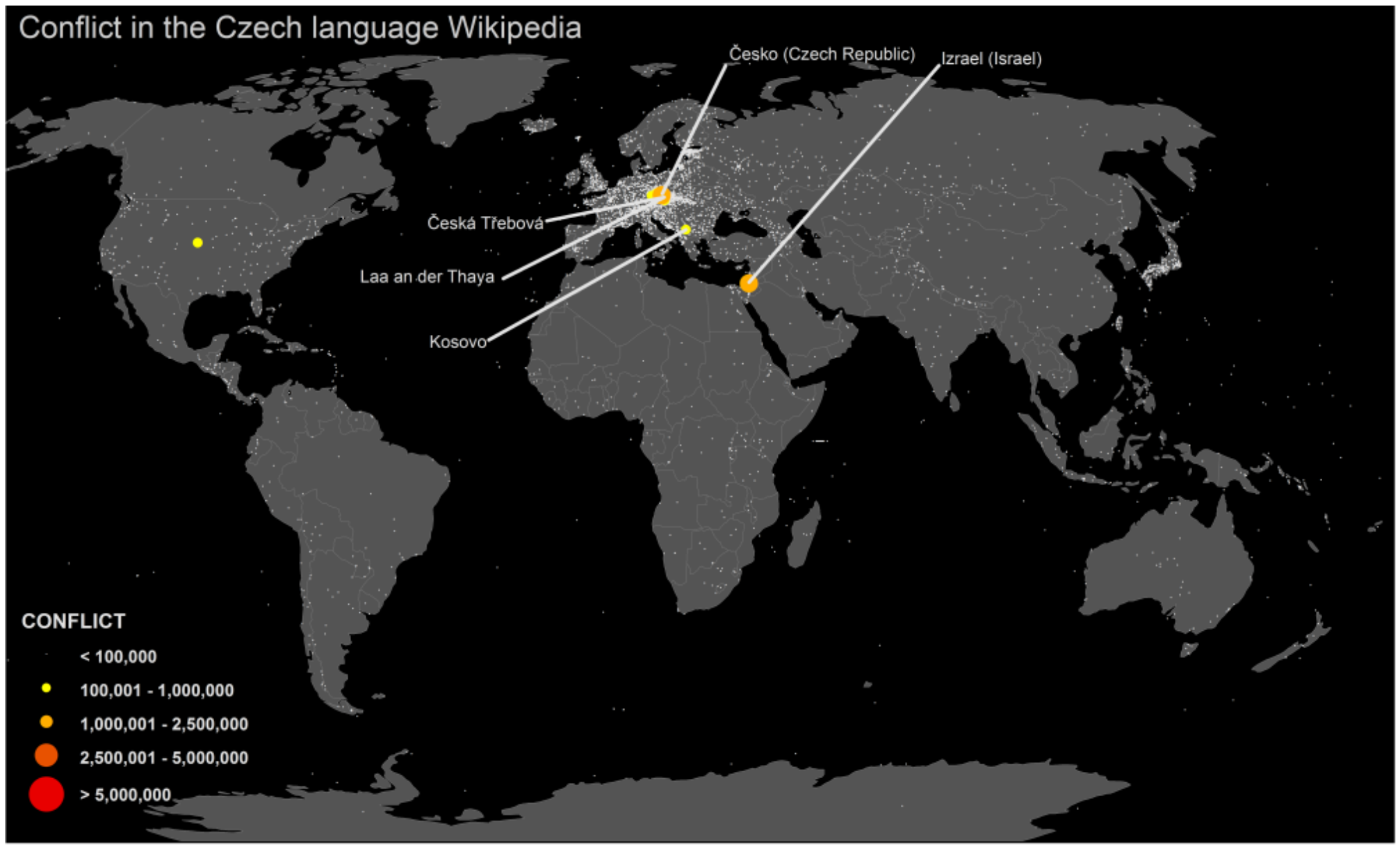

*Figure 10* Map of conflict in Czech edition of Wikipedia. Size of the dots is proportional to the controversy measure M.

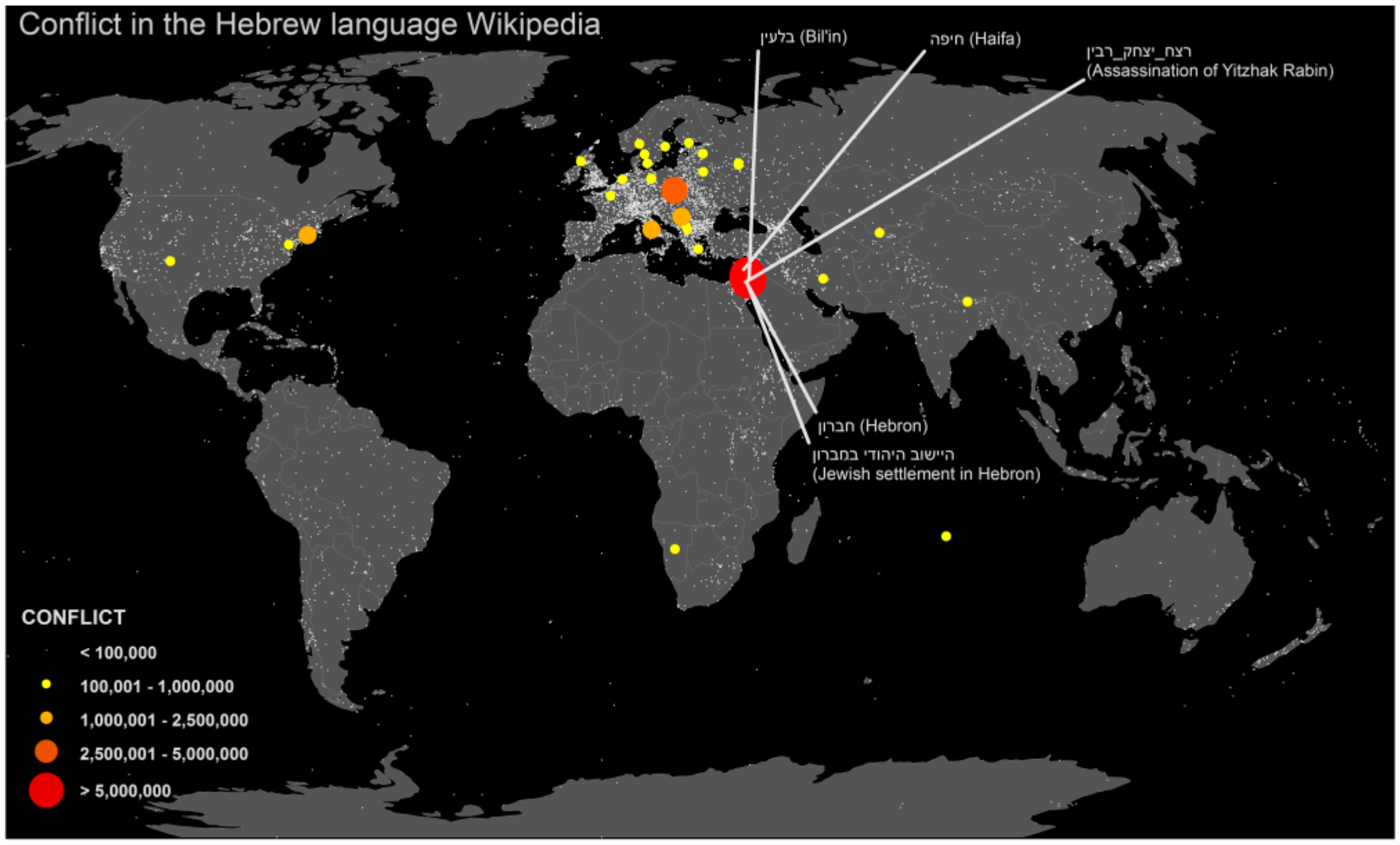

*Figure 11* Map of conflict in Hebrew edition of Wikipedia. Size of the dots is proportional to the controversy measure M.

Even when looking at large languages that are primarily spoken in more than one country, we are able to see that a significant amount of self-focus occurs (as can been seen in the maps of conflict in Arabic and Spanish in Figures 12 and 13). However, interestingly, the Middle East often seems to be the exception to this rule. The Spanish and Czech (as well as all languages in our sample apart from Hungarian, Romanian, Japanese, and Chinese) include articles in Israel as some of those characterised by the greatest amount of conflict.

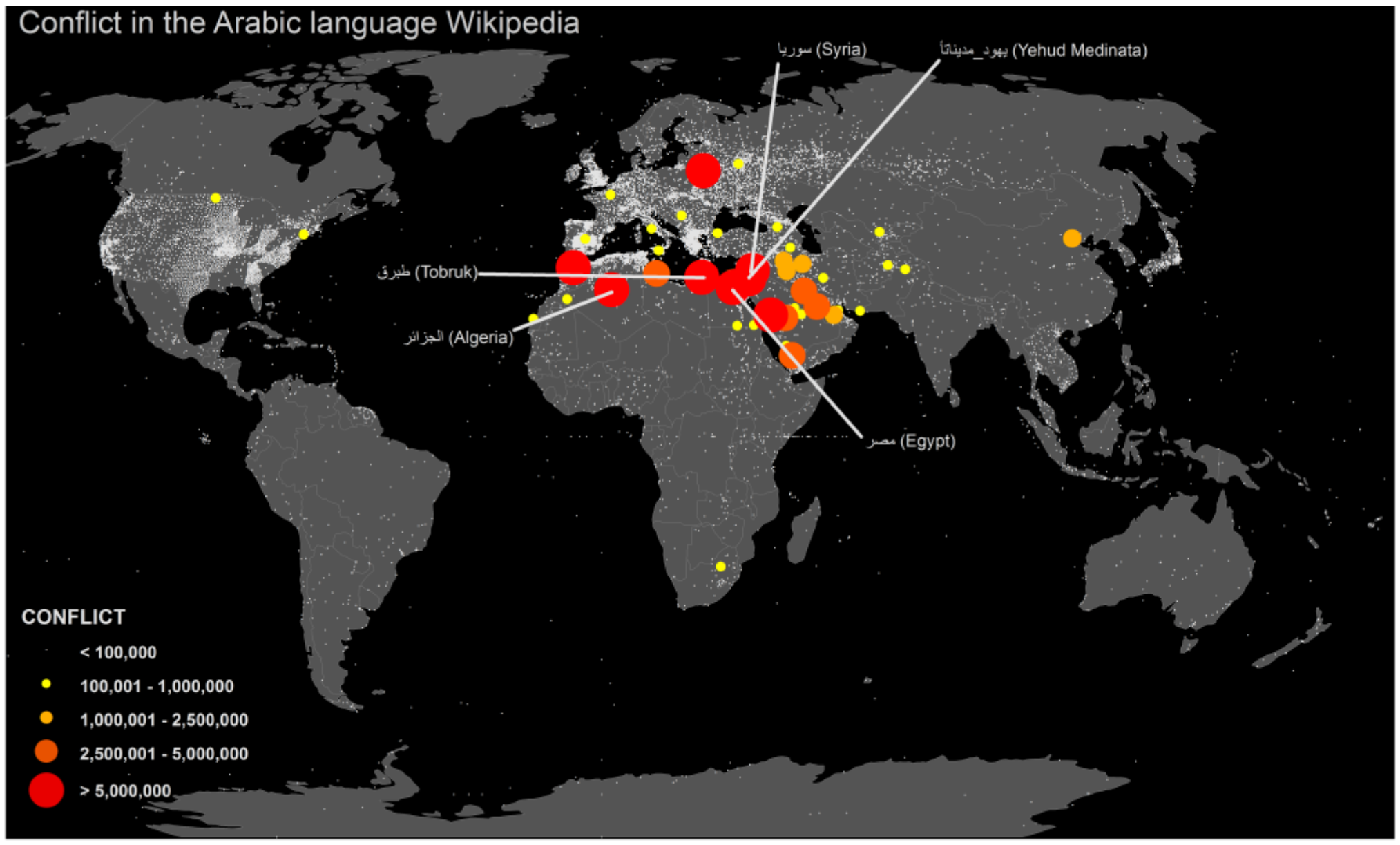

Figure 12 Map of conflict in Arabic edition of Wikipedia. Size of the dots is proportional to the controversy measure M.

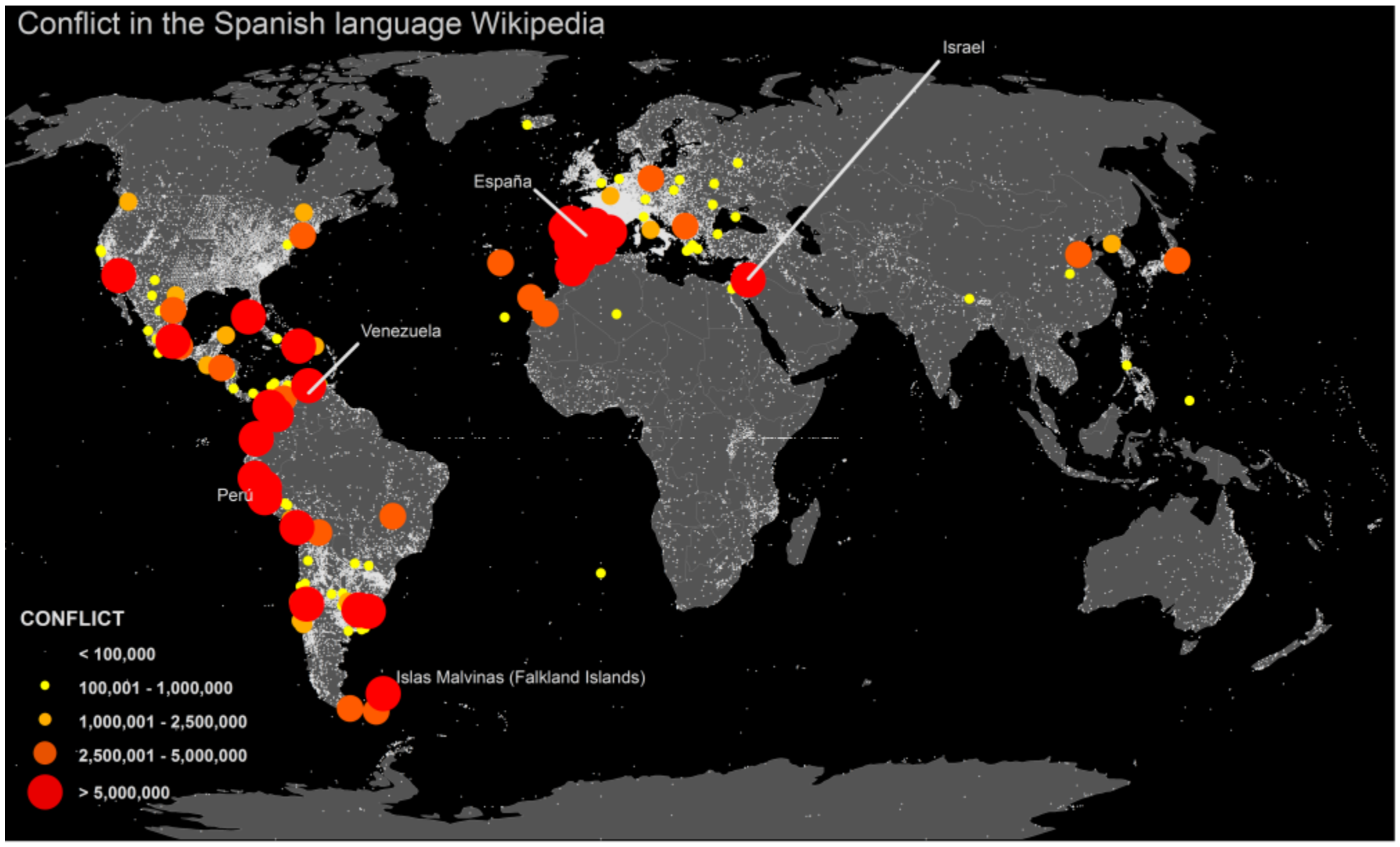

*Figure 13* Map of conflict in Spanish edition of Wikipedia. Size of the dots is proportional to the controversy measure M.

Also, worth noting is the fact that we see differences in the geographic topics that generate the most conflict. The articles in Japanese that generate the most conflict are not only all located in Japan, but are also all educational institutions. The Portuguese articles that generate the most conflict are similarly all located in Brazil (the world's largest Portuguese-speaking nation), with four out of the top five conflict scores being about football teams.

Within our sample, we actually only see the English, German, and French (shown in Figure 14) Wikipedias with a significant amount of diversity in the topics and patterns of conflict in geographic articles: indicating a less significant role that specific editors and arguments play in these larger encyclopaedias. More maps of different languages editions are available at http://www.zerogeography.net/2013/05/mapping-controversy-in-wikipedia.html.

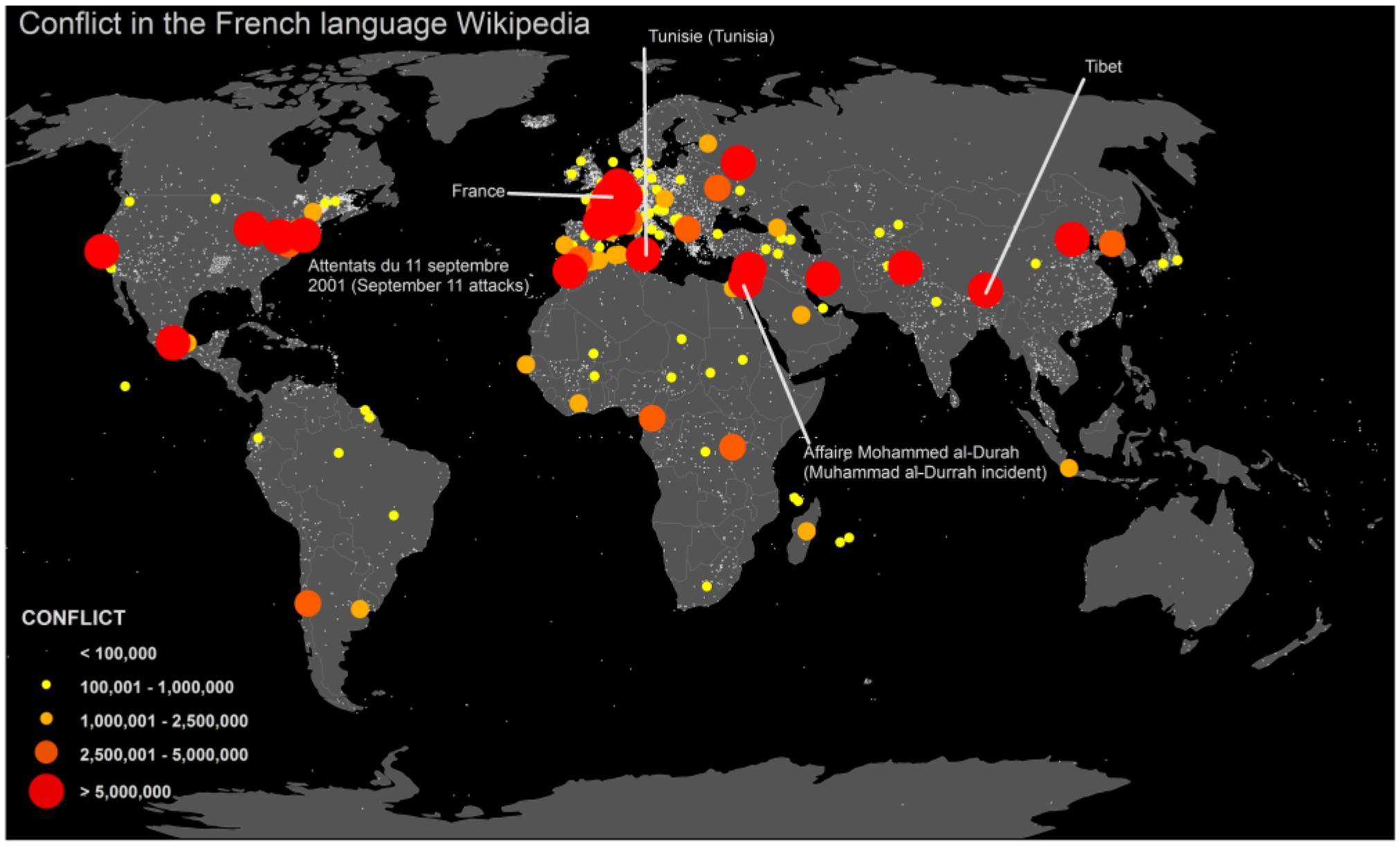

*Figure 14* Map of conflict in French edition of Wikipedia. Size of the dots is proportional to the controversy measure M.

## 4. Conclusion

In this chapter we have focused on two aspects of the conflict that occurs in different language versions of Wikipedia. First we explored the overlaps between the most contested articles in different languages. Second, we mapped out the geographical localities of the controversial articles based on their content.

The comparison of the contested articles in multiple languages has demonstrated that there are controversial topics, which are present in several regions and in Wikipedias in many different languages. In particular, the different languages are grouped into three sets: 1) English, German, French, Spanish; 2) Czech, Hungarian, Romanian; 3) Arabic, Persian, Hebrew. The articles *Israel*, *Adolf Hitler*, *The Holocaust* and *God* are highly contested in all the three language sets and the *Jesus*, *Islam*, and *Muhammad* articles are very highly contested in two of three sets of languages. Major religions and religious figures as well as articles related to Anti-Semitism and Israel are highly contested in multiple languages and cultures. A somewhat larger category is formed by the controversial articles which occur on a regional scale like those related to Eastern European history or local conflicts. However, somewhat surprisingly, most of the contested and controversial topics are language dependent.

The English Wikipedia, in particular, occupies a unique role. The language's status as a *lingua franca*, means that English Wikipedia ends up being edited by a broad community beyond simply that have the language as a mother tongue (Yasseri et al. 2012b, Circadian). As a result, it is expected that globally disputed themes are often represented in this Wikipedia. Already in the top-ten list of conflict articles we see such items as "Jesus", "Anarchism" or "Race and intelligence".

Within our sample, we actually only see that the English, German, and French Wikipedias have a significant amount of diversity in the topics and patterns of conflict in geographic articles. This probably indicates the less significant role that specific editors and arguments play in these larger encyclopaedias.

Ultimately by visualizing the controversy in Wikipedia, we're able to see both topics that appear to have cross-linguistic resonance (e.g. Arab-Israeli conflict), and those of more narrow interest such as the Islas Malvinas/Falkland islands article in the Spanish Wikipedia. The data presented here therefore offers a window into not just the topics and places that different language communities are interested in, but also the topics that seem worth fighting about.

This visualization supported research focused on a static picture as obtained from the statistics of conflict articles in different Wikipedias. The controversiality measure $M$ is, however, a dynamic quantity; it allows us to follow the temporal evolution of conflicts (Yasseri et al. 2012a Conflict, Török et al. 2013). It therefore remains a future task to combine the techniques used here with the study of those dynamic aspects. Furthermore, we see that this work could offer a useful base for more grounded qualitative and critical inquiry into the variable patterns of interest and controversy amongst different groups in the world's largest encyclopaedia.

## 6. Acknowledgment

We thank Robert Sumi, András Kornai, András Rung, Hoda Sepehri Rad for discussions and computational support.

Maps are created based on a dataset of geolocated Wikipedia articles prepared by Mark Graham, Bernie Hogan, and Ahmed Medhat.